\documentclass[12pt]{JHEP}

\usepackage{amsmath,epsfig}
\usepackage{amssymb,amsfonts}
\usepackage{latexsym}
\usepackage{longtable}

\relax

\def\be{\begin{equation}}
\def\ee{\end{equation}}
\def\bea{\begin{eqnarray}}
\def\eea{\end{eqnarray}}

\newcommand\fverb{\setbox\pippobox=\hbox\bgroup\verb}
\newcommand\fverbdo{\egroup\medskip\noindent%
                        \fbox{\unhbox\pippobox}\ }
\newcommand\fverbit{\egroup\item[\fbox{\unhbox\pippobox}]}

\newcommand{\bear}{\begin{eqnarray}}

\newcommand{\eear}{\end{eqnarray}}

\newbox\pippobox

\def\6{\partial}

\def\a{\alpha}
\def\nn{\nonumber}

\def\pa{\partial}

\def\e{\epsilon}
\def\m{\mu}
\def\n{\nu}

\def\s{\sigma}

\def\sp{\;\;\;,\;\;\;}

\def\sq
\def\a{\alpha}

\def\l{\lambda}

\def\hri#1#2{\href{http://arxiv.org/abs/#1}{[ArXiv:#1]#2}}
\def\hre#1#2{\href{http://arxiv.org/abs/#1/#2}{[ArXiv:#1/#2]}}

\def\na{\nabla}

\def\e{\epsilon}

\title{Hor\v ava-Lifshitz Cosmology}
\author{{\large Elias Kiritsis, Georgios Kofinas }
~\\
~\\
Department of Physics, University of Crete,
71003 Heraklion, Greece}

\preprint{}

\abstract{The cosmological equations suggested by the non-relativistic renormalizable  gravitational theory proposed by Ho\v rava
are considered. It is pointed out that the early universe cosmology has features that may give an alternative to inflation
and the theory may be able to escape singularities. }

\begin{document}

\section{Introduction and results}
\label{intro}
The UV completion of gravity has been an active arena of research in the past several decades.
A lot of understanding was gained by studying string theory and its gauge-theory incarnations.

Recently, a different field theory model for a UV complete theory of
gravity was proposed that is expected to be well defined and
complete in the UV \cite{hor2,hor3}, that we will call Hor\v ava-Lifshitz
gravity. The theory  does not have the full diffeomorphism
invariance of GR but only a subset (that is a form of local Galilean
invariance). This is the subset that is manifest in the ADM slicing
of standard GR.

Non-relativistic theories of gravity are motivated by attempts to
describe non-relativistic field theories via AdS/CFT. In particular,
in \cite{xliu} non-trivial bulk solutions were found that described
 backgrounds with non-relativistic scaling invariance.\footnote{Further progress along
 these lines was reported in \cite{t1,t2}.}
Instead of the AdS geometry exhibiting standard relativistic conformal invariance,
 a new geometry was found exhibiting Lifshitz scale invariance
\be
t\to \ell^z~t\sp x^i\to \ell~x^i
\ee
with $z\geq 1$. For $z=1$ this is the standard relativistic scale invariance.
Such solutions were found by modifying the standard UV AdS structure by the condensation of non-renormalizable operators.
The theory of gravitational fluctuations around such vacua is expected to exhibit a
similar breaking of relativistic invariance and a
modified (anisotropic) scaling symmetry.

Similar non-relativistic gauge theories can be defined \cite{hor1}. They have distinct critical dimensions.
They have a large-N limit at which such theories are expected to be dual to the solutions found in \cite{xliu}.

Non-relativistic scaling allows for many non-trivial scaling theories in dimensions $D>2$.
Indeed, when it comes to relativistic theories, two dimensions provides the richest class of scale invariant theories.
As we go up in dimension, the critical points become rare, and above 4 dimensions, no critical theory has a weakly coupled limit.
This can be changed with anisotropic scaling. For example, theories of scalars in  D spatial dimensions seem to have the same richness as two dimensional theories
($\sigma$-models). Lifshitz theories are therefore a new window into non-trivial critical physics in $D>2$ dimensions.

It is interesting that this picture allows a theory of gravitation that is scale-invariant in the UV, and where standard
general relativity with its higher symmetry
is an emerging theory in the IR. Moreover it seems that diffeomorphism invariance seems to be always an IR fixed point.

The introduction and study of non-relativistic gravitational theories and associated
 gauge theories was performed in \cite{hor1}-\cite{klu}.
 In particular, a theory with Lifshitz invariance corresponding to $z=3$ was
 argued to provide a ghost-free UV-renormalizable theory of non-relativistic gravity
 at least around flat space. This theory has two classically marginal couplings and
 several relevant operators corresponding to terms quadratic, linear
 and zeroth-order in the curvature of 3d space. The main marginal coupling is controlling the square of the
 Cotton tensor, while the second one, $\l$,  is controlling the contribution of
trace of the extrinsic curvature of space.

 At $\l={1\over 3}$ the theory exhibits an anisotropic Weyl symmetry. On the other hand, the value $\l=1$ is the one that
 is at play in standard GR.

 The presence of the Cotton tensor adds terms of order ${\cal O}(k^6)$ in the propagator
  and this improves the UV behavior rendering the theory renormalizable by power counting.
 This is similar to previous attempts to define a renormalizable theory of gravity by
  adding curvature square terms in order
 to improve the UV behavior \cite{tom}.
 The main difference is that previous work added higher time derivatives that resulted in a theory containing ghost excitations.
 Here, however, the time derivatives are kept at two at the expense of general coordinate invariance.

The two main and interesting properties of the theory are: (a) the
fact that it is UV renormalizable, (b) the effective speed of light
diverges in the UV. Further properties include a UV spectral
dimension  that is two \cite{hor4}, as well the appearance of
anisotropic Weyl invariance at $\l=1/3$.

Such interesting properties of the theory beg for a contact with
standard GR, that is a well-tested theory in the IR. Indeed, the
addition of relevant operators, notable that of the Einstein term,
as well as possibly a cosmological constant, drive the theory in the
IR to a variant of GR. It becomes exactly GR if $\l=1$. However,
$\lambda$ is a dimensionless coupling of the theory
 and therefore runs (logarithmically in the UV).

It has been argued in \cite{hor2} that $\lambda$ will eventually reach three possible IR fixed points: $\l=1$, $\l=1/3$ or $\l=\infty$.
Therefore the nature or the IR theory and its matching to GR depends crucially on the RG pattern.

The second issue is related to decoupling. Gravitational theories
with partly broken diffeomorphism invariance are known to have
peculiar patterns of decoupling, where intermediate scales appear
that are background-dependent \cite{vain}-\cite{Rub}. For example,
in generic massive gravity the strong coupling cutoff can be
described either in terms of the graviton mass $m_g$ ($\Lambda\sim
(m_g^4M_P)^{1\over 5}$), or in terms of the strength of the
background via the Vainshtein formula.
 Even in GR (and its quantum version as advocated in string theory) it was advocated, motivated by the black-hole information studies,  that
decoupling scales are background-dependent \cite{mathur}.

This peculiar pattern of non-decoupling in modified theories of
gravity is associated with the presence of non-linear instabilities
\cite{BD}, as well as with strong-coupling of some field theory
modes \cite{AGS}. The two problems were related in specific contexts
\cite{DR}. The strong coupling problem renders the associated
quantum theory obscure. Even in cases where the quantum theory is
well defined via a holographic correspondence, these awkward
features of modified gravity theories were shown to persist
\cite{KN}.

We now return to the novel properties of the theory and their potential implications.
The renormalizability of the theory implies that the fundamental theory
can be specified by a finite number of UV parameters
that can be in principle measured in the IR.

The fact that the speed of light is diverging in the UV opens up an
interesting possibility: that no inflation is needed at early times
in the evolution of the universe in order to solve the horizon
problem. The flatness problem maybe also potentially addressed as
spatial curvature is enhanced by the higher curvature terms. It is
therefore possible that a suppression of spatial curvatures may be
in operation at early times. Of course, a detailed study of the
cosmological perturbations will be needed in order to verify that
any substitute for inflation is in agreement with observational
data. Moreover, it is expected that the short distance structure of
perturbations will be drastically different from standard inflation.
At the end of this paper we will do a qualitative study of
cosmological perturbations and argue
 that indeed the previous expectations
can be realized in Ho\v rava-Lifshitz cosmology.

The UV structure of the theory  is also related with the possibility that curvature singularities are absent in
 the theory. Both cosmological and black-hole singularities are included
here. Indeed it is not unusual in higher-derivative theories of gravity that such singularities are absent
\cite{ar}. The important issue is that here, unlike many other higher-derivative examples, there is a rational
for having a finite number of such terms unlike the standard non-renormalizable terms in effective theories of gravity.

In this paper we couple the Ho\v rava-Lifshitz gravity with matter
and derive the classical equations of motion, allowing a more
general class of couplings, not restricted by detailed balance. We
further make a cosmological ansatz and derive the analogue of the
Friedmann equation. As expected, most of the non-relativistic
structure does not enter in this equation. There are, however,
contributions from the quadratic curvature terms that dominate in
the UV (when the universe is small) and provide a bounce. This is in
qualitative agreement with our expectations above.

We also describe
the qualitative structure of the equations for cosmological
perturbations, and conclude
 that the theory can lead to an alternative of inflation in the UV regime.
It can solve the horizon and flatness problems, can generate scale-invariant superhorizons perturbations and does not
need graceful exit nor reheating.

\section{The Ho\v rava-Lifshitz gravity theory}

We review here the Ho\v rava-Lifshitz gravitational theory as was formulated in \cite{hor3}.

The dynamical variables are $N,N_i,g_{ij}$, with scaling dimension
zero, except $N_i$ that has scaling dimension 2. This is similar to
the ADM decomposition of the metric in standard general relativity,
where the metric is written as
\be ds^2=-N^2
~dt^2+g_{ij}(dx^i+N^idt)(dx^j+N^jdt)\sp N_i=g_{ij}N^j. \label{1}
\ee
The scaling transformation of the coordinates is now modified to \be
t\to \ell^3~t\sp x^i\to \ell~x^i, \label{15}\ee under which $g_{ij}$
and $N$ are invariant, while $N^i$ scales as $N^i\to \ell^{-2} N_i$.

The kinetic terms are given by \be S_K={2\over \kappa^2}\int dtd^3x
\sqrt{g}N\left(K_{ij}K^{ij}-\l K^2\right)\sp K=g^{ij}K_{ij}\sp
K^{ij}=g^{ik}g^{jl}K_{kl} \label{2}\ee in terms of the extrinsic
curvature \be K_{ij}={1\over 2N}(\dot
g_{ij}-\nabla_{i}N_j-\nabla_jN_i), \label{3}\ee with covariant
derivatives defined with respect to the spatial metric $g_{ij}$.

The potential is given in the ``detailed-balance" form
\be
S_{V}=-{\kappa^2\over 8}\int dt d^3x\sqrt{g}N~E^{ij}{\cal G}_{ij;kl}E^{kl}
=\int dt d^3x\sqrt{g}N~\left[-{\kappa^2\over 2w^4}C_{ij}C^{ij}+\right.
\label{4}\ee
$$
\left. +{\kappa^2\mu\over 2w^2} {\e^{ijk}\over
\sqrt{g}}R_{il}\na_j{R^l}_{k}-{\kappa^2\mu^2\over
8}R_{ij}R^{ij}+{\kappa^2\mu^2\over 8(1-3\l)} \left({1-4\l\over
4}R^2+\Lambda~R-3\Lambda^2\right)\right],
$$
where the supermetric  ${\cal G}^{ij;kl}$ depends on an a priori arbitrary parameter $\l$
\be
 {\cal G}^{ij;kl}={1\over 2}\left(g^{ik}g^{jl}+g^{il}g^{jk}\right)-\lambda g^{ij}g^{kl}
 \sp  {\cal G}_{ij;kl}={1\over 2}\left(g_{ik}g_{jl}+g_{il}g_{jk}\right)+{\lambda\over 1-3\l} g_{ij}g_{kl}
\label{5}\ee
\be {\cal G}^{ij;kl}{\cal G}_{kl;mn}={1\over
2}({\delta^i}_{m}{\delta^j}_{n}+{\delta^i}_{n}{\delta^j}_{m}),
\label{6}\ee and the $E$ tensors are given by \be E^{ij}={2\over
w^2}C^{ij}-{\mu}\left(R^{ij}-{1\over 2}Rg^{ij}+\Lambda
g^{ij}\right), \label{7}
\ee
 where
 \be C^{ij}={\e^{ikl}\over
\sqrt{g}}\na_k\left({R^j}_{l}-{1\over 4}R{\delta^{j}}_{l}\right)
\label{8}\ee
is the Cotton tensor (conserved and traceless,
vanishing for conformally flat metrics). $E^{ij}$ can be  derived
from an action as \be
 E^{ij}={1\over \sqrt{g}}{\delta
W\over \delta g_{ij}}\sp W={1\over w^2}\int \omega_3(\Gamma)+\mu\int
d^3x\sqrt{g}(R-2\Lambda), \label{9}\ee
 with
 \be
{\omega_3}(\Gamma)=Tr\left(\Gamma\wedge d\Gamma+{2\over
3}\Gamma\wedge \Gamma\wedge \Gamma\right) = {\e^{ikl}\over
\sqrt{g}}\left({\Gamma^m}_{il}\partial_{j} {\Gamma^l}_{km}+{2\over
3}{\Gamma^n}_{il} {\Gamma^l}_{jm}{\Gamma^m}_{kn}\right),
\label{10}\ee and ${\Gamma^i}_{jk}$ are the Christoffel symbols of
$g_{ij}.$ The full action can be rewritten as \be S=S_K+S_V=\int
dtd^3x\sqrt{g}N\left[{2\over \kappa^2}(K_{ij}K^{ij}-\l
K^2)-{\kappa^2\over 2w^4}C_{ij}C^{ij}+ {\kappa^2\mu\over 2w^2}
{\e^{ijk}\over \sqrt{g}}R_{il}\na_j{R^l}_{k}+\right. \label{11}\ee
$$
\left.-{\kappa^2\mu^2\over 8}R_{ij}R^{ij}+{\kappa^2\mu^2\over
8(1-3\l)}\left({1-4\l\over 4}R^2+\Lambda~R-3\Lambda^2\right)
\right],
$$
with $ {\cal E}^{ijk}={\e^{ijk}\over \sqrt{g}}$ the standard generally
covariant antisymmetric tensor. $\e^{123}$ is defined to be 1, and
other components are obtained by antisymmetry. Indices are raised
and lowered with the metric $g_{ij}$. Therefore, ${\cal E}^{ijk}=(\pm 1)
/\sqrt{g}$.

We will rewrite the action (\ref{11}) as \be S\!=\!\!\int\!
dtd^3x\sqrt{g}N\!\!\left[\alpha (K_{ij}K^{ij}\!-\!\l K^2)\!+\!\beta
C_{ij}C^{ij}\!+\!\gamma {\cal E}^{ijk}R_{il}\na_j{R^l}_{k}
\!+\!\zeta R_{ij}R^{ij}\!+\!\eta R^2\!+\!\xi
R\!+\!\sigma\!\right]\!, \label{12}\ee with \be\!\!\!\!\!\!\!\!\!\!
\alpha\!=\! \frac{2}{\kappa^{2}}\sp \beta\!=\!
-\frac{\kappa^{2}}{2w^{4}}\sp \gamma\!=\!
\frac{\kappa^{2}\mu}{2w^{2}}\sp
\zeta\!=\!-\frac{\kappa^{2}\mu^{2}}{8} \label{13}\nn\ee \be
\eta\!=\!
\frac{\kappa^{2}\mu^{2}}{8(1\!-\!3\lambda)}\frac{1\!-\!4\lambda}{4}\sp
\xi\!=\! \frac{\kappa^{2}\mu^{2}}{8(1\!-\!3\lambda)}\Lambda\sp
\sigma\!=\!
\frac{\kappa^{2}\mu^{2}}{8(1\!-\!3\lambda)}(-3\Lambda^{2}).
\label{14}\ee
 The reason is that we eventually would like to explore more general
 possibilities compared to the detailed-balance ansatz in  (\ref{4}).

 Indeed, the action in (\ref{12}) is the most general action that is renormalizable around flat space
 and contains apart from the Cotton tensor, quadratic terms in the curvatures.
 The most general renormalizable action might also contain terms cubic in the curvatures.
 The term quadratic in the Cotton tensor is marginal, while the rest of the terms are relevant under the scaling
 (\ref{15}).

The action in (\ref{12}) is invariant under a restricted class of
diffeomorphisms \be t'=h(t)\sp x'^i=h^i(t,\vec x). \label{16}\ee The
transformation of the metric under infinitesimal diffeomorphisms is
\be \delta
g_{ij}=\pa_i\e^kg_{jk}+\pa_j\e^kg_{ik}+\e^k\pa_kg_{ij}+f\dot g_{ij}
\ee \be \delta N_i=\pa_i\e^jN_j+\pa_j\e^jN_i+\dot \e^j g_{ij}+\dot f
N_i+f\dot N_i\sp \delta N =\e^j\pa_j N+\dot f N+f{\dot N}. \ee

\subsection{The IR limit}

In the IR the action simplifies \be S\to S_E=\int dtd^3x
\sqrt{g}N\left[\alpha(K_{ij}K^{ij}- \lambda K^2)+\xi
R+\sigma\right]. \label{17}
\ee
Defining $x^0=ct$, choosing
$\lambda=1$ and
\be c=\sqrt{\xi\over \alpha}\sp 16\pi
G=\frac{1}{\sqrt{\alpha\xi}}\sp \Lambda_E=-{\sigma\over 2\xi},
\label{18}
\ee
the action is that of Einstein
\be S_E={1\over 16\pi
G}\int d^4x \sqrt{\tilde g}\left[K_{ij}K^{ij}-
K^2+R-2\Lambda_E\right]={1\over 16\pi G}\int d^4x \sqrt{\tilde g}
\left[\tilde R-2\Lambda_E\right]. \label{19}
\ee
The full space-time
metric $\tilde g_{\m\n}$ is \be \tilde g_{00}=-N^2+N_{i}g^{ij}N_j\sp
\tilde g_{0i}=N_i\sp \tilde g_{ij}=g_{ij}\sp \det[\tilde g]=\det[g]
N^2, \label{20}\ee while the inverse metric $\tilde g^{\m\n}$ \be
\tilde g^{00}=-{1\over N^2}\sp \tilde g^{0i}={N^i\over N^2}\sp
\tilde g^{ij}=g^{ij}-{N^iN^j\over N^2}. \label{21}\ee

For the detailed-balance ansatz it is \be c={\mu\kappa^2\over
4}\sqrt{\Lambda\over 1-3\lambda}\sp G={\kappa^2\over 32\pi c}.
\label{22}\ee

The previous analysis was classical, but in the quantum theory the
situation is more involved. There are two additional issues that
obscure the low energy picture, that were discussed in the first
section. These include the fact that $\l$ is a classically marginal
coupling that runs in the quantum theory.\footnote{We are indebted
to P. Ho\v rava for illuminating comments.} The second involves the
issue of symmetries, namely the
 enhancement of the non-relativistic local invariance to full diffeomorphism invariance. As argued in the fist section, this is a tricky issue
 that can be at the source of many problems and subtleties.

It is   therefore an open problem that needs further investigation to what extend non-relativistic gravity asymptotes to GR
in the IR, and at what scale the first corrections appear.

\section{Non-relativistic Matter}

\subsection{The scalar action}

To couple matter to non-relativistic gravity we follow the same
strategy:
 write a quadratic kinetic term with the right symmetries and add a potential.
We will not insist here on detailed-balance.

We will start from the simplest case of a scalar. In the
non-relativistic case a general  action can be
written as
\be S_{\rm nr}=\int d^3xdt \sqrt{g}N\left[{1\over
N^2}(\dot \Phi-N^i\partial_i\Phi)^2-F(\partial_i\Phi,\Phi)\right],
\label{23}
\ee
where the potential $F$ can in principle contain
arbitrary combinations of $\Phi$ and derivatives.

For the kinetic term to have marginal scaling, $\Phi$ must be
dimensionless, i.e. it should transform as $\Phi\to \Phi$ under a
non-relativistic conformal transformation. Therefore, UV
renormalizability indicates that $F$ should contain up to six
derivatives, but otherwise arbitrary powers of the scalar field. We
rewrite the action by taking $F$ to be  a function of $\xi_n
=\Phi\square^n\Phi $ and $\Phi$ where $\square$ is the
three-dimensional Laplacian of the metric $g_{ij}$
\be S_{\rm
nr}=\int d^3xdt \sqrt{g}N\left[{1\over N^2}(\dot
\Phi-N^i\partial_i\Phi)^2-F[\xi_1,\xi_2,\cdots,\Phi]\right].
\label{24}
\ee
For UV renormalizability, $F$ must contain at most
six derivatives
\be
F[\xi,\Phi]=F_{0}(\Phi)+F_{1}(\Phi)~\xi_1+F_{11}(\Phi)\xi_1^2+F_{111}
(\Phi)\xi_1^3+F_{2}(\Phi)\xi_2+F_{21}(\Phi)\xi_2\xi_1+F_3(\Phi)\xi_3.
\label{25}
\ee
In the UV it is the terms that involve $\xi_1^3$,
$\xi_1\xi_2$ and $\xi_3$ that dominate. In the IR, the theory is
dominated by the $F_0(\Phi)$ and the $F_1(\Phi)$ terms. By a field
redefinition we can make $F_1$ to be a constant (the IR speed of
light), and the theory
 reduces to a scalar field with a potential generated by $F_0(\Phi)$.

The equations of motion stemming from the variation of the action
are
\be {1\over \sqrt{g}}\partial_t\left[{\sqrt{g}\over N}(\dot
\Phi-N^i\partial_i\Phi)\right]-\na_j\left[{N^j\over N}(\dot
\Phi-N^i\partial_i\Phi)\right] +{N\over 2}{\delta F\over \delta
\Phi}=\label{26}
\ee
$$
=-{1\over 2}\sum_{n}\left[\square^n\left(N\Phi{\delta F\over \delta
\xi_n}\right)+N{\delta F\over \delta \xi_n}\square^n\Phi\right].
$$

We will also need the variations of the effective action with
respect to the metric
\be {N^2\over \sqrt{g}}{\delta S\over
\delta N}=-(\dot \Phi-N^i\partial_i\Phi)^2-N^2F \label{27}\sp {N\over
\sqrt{g}}{\delta S\over \delta N^i}=-2\pa_i\Phi(\dot
\Phi-N^j\partial_j\Phi) \ee
\be {1\over \sqrt{g}N}{\delta S\over
\delta g^{ij}}=-\sum_n{\delta F\over \delta
\xi_n}\Phi{\delta \square^n\over \delta g^{ij}}\Phi+{1\over 2}g_{ij} \left[-{1\over N^2}(\dot
\Phi-N^i\partial_i\Phi)^2+F[\xi,\Phi]\right].
\label{28}\ee

\subsection{The vector action}

We consider now  a (massive) vector. A general non-relativistic
action where the potential depends only on the gauge field, field
strength and first derivatives can be written as follows
\be S_{nr}={1\over 4g^2}\int
d^3xdt \sqrt{g}N\left[{2\over
N^2}g^{ij}(F_{0i}-N^kF_{ki})(F_{0j}-N^lF_{lj})+\right.\label{29}\ee
$$
\left.{M^2\over N^2}(A_0-N^iA_i)(A_0-N^jA_j)
-G[A_i]\right],
$$
where as usual $F_{0i}=\partial_t A_i-\partial_iA_0$, $F_{ij}=
\partial_i A_j-\partial_jA_i$ is the (abelian) field strength. The
scaling dimension of $A_i$ is zero, while that of $A_0$ is 2. The
function $G$ must be of the form
\be G=G[\zeta_i,g^{ij}A_iA_j],
\ee
where
\be
\zeta_1=B_iB^i\sp \zeta_2=\nabla_iB_j\nabla^i B^j
 \ee
\be
\zeta_3=\nabla_iB_j\nabla^i B^k \nabla^jB_k\sp
\zeta_4=\nabla_i\nabla_jB_k\nabla^i\nabla^j B^k,
\ee
and we
introduced the magnetic field
\be B_i={1\over
2}{{\epsilon_{i}}^{jk}\over \sqrt{g}}F_{jk}\sp
F_{ij}={{\epsilon_{ij}}^k\over {\sqrt{g}}}B_{k}\sp \nabla^iB_i=0.
\ee
Note that there are terms related by integration by parts to the
ones introduced above but they produce no new effects in backgrounds
with $N$ space-independent. We have also included terms that will
result to a renormalizable action in the UV. In particular, for
renormalizability in the UV, the function G must be of the form
\be
G=a_0+a_1\zeta_1+a_2\zeta_1^2+a_3\zeta_1^3+a_4\zeta_2+a_5\zeta_1\zeta_2+a_6\zeta_3+a_7\zeta_4,
\ee
 where the coefficients $a_i$ can be arbitrary functions of
$A_iA^i$. In particular, the terms linear in $\zeta_1$ and $\zeta_4$
provide derivative contributions to the vector propagator. In
particular, the $\zeta_4$ term changes drastically the vector
interaction in the UV.

Renormalizability constraints the matter actions to contain in the
UV a finite number of terms, although this number is larger than
associated relativistic actions. The coefficients of such terms will
be constrained by experimental limits on Lorentz-invariance
violations. Moreover, quantum effects will link relativistic
invariance violations between the gravitational and the matter
sectors. Investigating such constraints quantitatively is an
interesting and important problem.

\section{The classical equations of motion\label{eq}}

We now add the action of matter \be S_{M}=\int d^3xdt
\sqrt{g}N~{\cal L}_{\rm matter}(N,N_i,g_{ij}) \label{30}\ee to the
gravitational action (\ref{12}) and we will vary with respect to the
gravitational fields to obtain the equations of motion.

The equation obtained by varying N  is
\be
-\alpha\left(K_{ij}K^{ij}-\l K^2\right)+\beta C_{ij}C^{ij}+\gamma
{\cal E}^{ijk}R_{il}\na_j{R^l}_{k} +\zeta R_{ij}R^{ij}+\eta R^2+\xi
R+\sigma=J_N, \label{31}
\ee
with
\be J_N=-{\cal L}_{\rm
matter}-N{\delta {\cal L}_{\rm matter}\over \delta N}.
\ee

The equation  obtained by varying  $N_i$  is
\be
2\alpha(\na_jK^{ji}-\l\na^i K)+N{\delta {\cal L}_{\rm matter}\over
\delta N_i}=0. \label{32}
\ee

Finally, the equation of motion obtained by varying $g_{ij}$ is more
voluminous

\bea &&\!\!\!\!\!\!\!\!\frac{1}{2}
\Big[({\cal E}^{mk\ell}Q_{mi})_{;kj\ell}\!+\!({\cal E}^{mk\ell}Q^{n}_{m})_{;kin}g_{j\ell}\!-\!
({\cal E}^{mk\ell}Q_{mi})_{;kn}^{\,\,\,\,\,\,\,;n}g_{j\ell}\!-\!({\cal E}^{mk\ell}Q_{mi})_{;k}R_{j\ell}\nn\\
&&\!\!\!\!\!\!\!\!\!\!\!
-({\cal E}^{mk\ell}Q_{mi}R^{n}_{k})_{;n}g_{j\ell}\!+\!({\cal E}^{mk\ell}Q^{n}_{m}R_{ki})_{;n}g_{j\ell}\!+\!
\frac{1}{2}({\cal E}^{mk\ell}R^{n}_{\,\,\,pk\ell}Q^{p}_{m})_{;n}g_{ij}\!-\!Q_{k\ell}C^{k\ell}g_{ij}\!+\nn\\
&&\!\!\!\!\!\!\!\!\!\!\!
{\cal E}^{mk\ell}Q_{mi}R_{j\ell;k}\Big]+\Box[N(2\eta
R\!+\!\xi)]g_{ij}\!+\!N(2\eta R\!+\!\xi)R_{ij}\!+\!2N(\zeta
R_{ik}R^{k}_{j}\!-\!\beta C_{ik}C^{k}_{j})\!\nn\\
&&\!\!\!\!\!\!\!\!\!\!\!-[N(2\eta R\!+\!\xi)]_{;ij}\!+\!\Box[N(\zeta
R_{ij}\!+\!\frac{\gamma}{2}C_{ij})]-2[N(\zeta
R_{ik}\!+\!\frac{\gamma}{2}C_{ik})]_{;j}^{\,\,\,\,;k}\!+\![N(\zeta
R^{k\ell}\!+\!\frac{\gamma}{2}C^{k\ell})]_{;k\ell}g_{ij}\!\nn\\
&&\!\!\!\!\!\!\!\!\!-\frac{N}{2}(\beta C_{k\ell}C^{k\ell}\!+\!\gamma
R_{k\ell}C^{k\ell}\!+\!\zeta R_{k\ell}R^{k\ell}\!+\!\eta
R^{2}\!+\!\xi
R\!+\!\sigma)g_{ij}\!+2\alpha\!N(K_{ik}K^{k}_{j}\!-\!\lambda
KK_{ij})\!\nn\\
&& -\frac{\alpha N}{2}(K_{k\ell}K^{k\ell}\!-\!\lambda
K^{2})g_{ij}\!+\!\frac{\alpha}{\sqrt{\!g}}g_{ik}g_{j\ell}{\partial\over
\partial t}[\sqrt{\!g}(K^{k\ell}\!-\!\lambda
Kg^{k\ell})]+\alpha[(K_{ik}\!-\!\lambda K g_{ik})N_{j}]^{;k}\!\nn\\
&& +\alpha[(K_{jk}\!-\!\lambda K
g_{jk})N_{i}]^{;k}\!-\!\alpha[(K_{ij}\!-\!\lambda K
g_{ij})N_{k}]^{;k}+(i\leftrightarrow j)=-2N{\delta {\cal L}_{\rm
matter}\over \delta g^{ij}}, \label{33}\eea \noindent where
${\cal E}^{ijk}$ was defined below (\ref{11}), $;$ stands for covariant
differentiation with respect to the metric $g_{ij}$,  and \be
Q_{ij}\equiv N(\gamma R_{ij}\!+\!2\beta C_{ij}). \label{34}\ee

\section{Cosmological metrics}
We will now consider solutions to the non-relativistic gravity
equations of the previous section which are of cosmological nature.
This in particular implies that they will be homogeneous and
isotropic. The associated ansatz is \be N\to N(t)\sp N_i\to 0\sp
g_{ij}=a^2(t)\gamma_{ij}, \label{35}\ee where $\gamma_{ij}$ is a
maximally symmetric constant curvature metric \be
R^{\gamma}_{ij}=2k\gamma_{ij}\sp R^{\gamma}=6k. \label{36}\ee We
will also assume scalar matter described by the action (\ref{24}).
The cosmological ansatz for the scalar is $\Phi\to \Phi(t)$.

There are many simplifications occurring for cosmological backgrounds.
First and foremost, as $\partial_i\Phi=0$ the non-linear function $F$
in the scalar action reduces effectively to a potential.
Therefore, for cosmological solutions the matter action behaves as
the relativistic one. This will not be however the case for perturbations.

In the gravitational sector also, spatial covariant derivatives
mostly vanish (only their commutator may be non-zero some times as
it gives further powers of the spatial curvature). Moreover, the
Cotton tensor vanishes. Therefore, there are simplifications,
however, we will obtain terms quadratic in curvatures that will be
the important novelty of the non-relativistic cosmological
equations.

The $N$-equation becomes
\be 3\alpha (3\l-1)H^2+\sigma+{6k\xi\over
a^2}+{12k^2(\zeta+3\eta)\over a^4}=-{1\over \sqrt{g}}{\delta
S_M\over \delta N}= {\dot \Phi^2\over N^2}+V(\Phi), \label{37}
\ee
with
\be H\equiv {\dot a\over Na}\sp V(\Phi)\equiv F(0,\Phi),
\label{39}
\ee

while the $N_i$ equations are trivial. The $g_{ij}$
equation gives \be 2\alpha(3\l-1)\left[\dot H+{3\over
2}H^2\right]+\left[\sigma+ {2k\xi\over a^2}-{4k^2(\zeta+3\eta)\over
a^4}\right]={2g^{ij}\over 3N\sqrt{g}}{\delta S_M\over \delta
g^{ij}}=-{\dot \Phi^2\over N^2}+V(\Phi), \label{40}\ee with \be \dot
H\equiv {1\over N}\partial_t\left({\dot a\over aN}\right).
\label{41}\ee Finally, the scalar equation becomes \be {1\over
N}\partial_t\left({\dot \Phi\over N}\right)+3H{\dot\Phi\over
N}+{V'\over 2}=0. \label{42}\ee We may use the residual invariance
under time reparametrizations to set $N=1$ below. As expected,
equation (\ref{40}) follows from (\ref{37}) and (\ref{42}).

The equations can be generalized to arbitrary matter using density
and pressure
\be 3\alpha (3\l-1)H^2=\rho-\sigma-{6k\xi\over
a^2}-{12k^2(\zeta+3\eta)\over a^4}\sp \dot\rho+3H(\rho+p)=0
\label{38}
\ee
 with
\be \rho=-{1\over \sqrt{g}}{\delta S_M\over
\delta N}\sp -{2\over N\sqrt{g}}{\delta S_M\over \delta
g^{ij}}=p~g_{ij}. \label{43}
\ee

A few comments are in order concerning the non-relativistic
Friedmann equation (\ref{37}). Its main differences compared to GR
with a cosmological constant are two. The first is the presence of
the parameter $\l$ that modifies the nature of the equation
depending on whether it is larger or small than 1/3. The second is
the presence of the $1/a^4$ term that resembles the ``mirage
radiation" term of Randall-Sundrum cosmology \cite{rsc}. This is not
an accident since Randall-Sundrum cosmology can be mapped via
holography to a four-dimensional problem with curvature square terms
\cite{holo}. However, unlike the RS case, this term here is related
to spatial curvature.
 In particular, it is absent in a spatially flat universe.

In the regime $\l>{1\over 3}$, both $\alpha$ and $\xi$ must be
positive for the IR field theory to be close to GR. In principle the
cosmological constant $\s$ can  be non-zero and can have either
sign. In the Ho\v rava formulation \cite{hor3}, it is positive
because of detailed balance, leading to anti-deSitter vacua.

The sign of the higher-derivative couplings like $\zeta,\eta$ can
also a priori have either sign. It has been argued in several
contexts based on unitarity that they must be positive. A quick way
to see the signals is to look at the dispersion relation, that after
rescaling will look like $\omega^2-\vec k^{\,2}+(\vec k^{\,2})^2$,
where we kept the curvature square terms but have ignored the Cotton
tensor. This dispersion relation has no solutions above a certain
momentum, signaling pathological high-momentum behavior.
 However here, this is not a problem once the Cotton tensor terms are present as they dominate the high-energy behavior.
 This a priori allows the possibility that the mirage radiation term is negative as it happens in other well studied cases
 associated to D-brane backgrounds
 \cite{mirage}. In the detailed-balance formulation, such terms are positive, leading to an effective negative contribution to the energy.
 In this case, their presence leads to a bouncing behavior when the universe is small. This statement is independent of the sign of the
 cosmological constant
 and standard curvature term.
 Moreover, it allows for the possibility of avoiding the initial singularity in this context.

 The case $\l<{1\over 3}$ is very different. As obvious from the equations this is equivalent to flipping the sign of the energy.
 Upon this flip, similar remarks hold as in the previous case.

 Finally, in the case with $\lambda={1\over 3}$ a new conformal (anisotropic) invariance develops \cite{hor3}.
Although the theory is scale invariant in the UV for any $\lambda$, in this case this scale invariance becomes a local symmetry.
 In this case, the extra invariance allows scaling away the scale factor to a constant which is non-dynamical.
 This is compatible with dropping the $H^2$ term in the Friedmann equation (\ref{38}).

The most general renormalizable theory contains also cubic terms in
the curvature. In three dimensions the Riemann tensor can be written
in terms of the Ricci tensor and Ricci scalar. Therefore the most
general such terms are \be S_{R^3}=\int
dt~d^3x\sqrt{g}N\left[\delta_1 ~R_{ij}{R^{i}}_kR^{jk}+\delta_2~
R_{ij}R^{ij}R+\delta_3 ~R^3+\delta_4~R\square R+\delta_5~R_{ij}\square R^{ij} \right]. \label{r3}\ee Unlike curvature
square terms that contribute to the propagator, cubic terms
contribute to the cubic graviton interaction vertex. In general, we
would expect them to be generated by renormalization even if they
are absent in the tree-level action. This however needs further
investigation.

Their contribution to the Friedmann equation (\ref{37}) is to add
the term \be (3\delta_1+3^2\delta_2+3^3\delta_3){(2k)^3\over a^6}
\label{r4}\ee to the left-hand side. In this case, the effect of
non-zero spatial curvature is even  stronger when the universe is
small. It has the same behavior as that of a massless scalar.

\subsection{Remarks on cosmological perturbations}

As mentioned in the introduction the Ho\v rava-Lifshitz theory of
gravitation is expected to have interesting and potentially welcome
consequences for early universe cosmology and in particular for the
inflationary era. In particular, we expect that no exponential
expansion of the universe will be needed to solve the horizon and
flatness problems, and to generate scale invariant perturbations for
the early universe. Let us call the cosmological era where the
six-derivative terms dominate the dynamics, the Ho\v rava-Lifshitz
era. Here, the speed of light is (almost) infinite and this allows
for a solution of the horizon problem. Indeed, we will see below
that the scale invariance of the theory in the UV  generically
creates large scale perturbations that are super-horizon without the
need for inflation.

As we have already seen in the previous section, the effects of spatial curvature are amplified in the early universe.
Keeping them of the same order as other matter densities will suppress curvature contributions in later times,
providing thus the possibility of solving the flatness problem.
Finally, the theory is scale invariant and we expect therefore the
perturbation that will be generated in that era to be scale invariant.
The reason is that in this theory, at early times, the scalar fields
 have zero dimension, therefore, their fluctuations are expected to be scale independent.
This is in contrast with standard relativistic scalars that have
dimension one, and therefore their fluctuations scale as the energy
(or Hubble parameter).

Deriving the equations for cosmological perturbations in this
context is a formidable task. We can, however, anticipate some of
the structure of perturbation equations. Consider for example scalar
perturbations of an arbitrary fluid, whose energy density is
parametrized by some gauge invariant variable $\varepsilon\sim
{\delta \rho\over \rho}$. In standard GR such perturbations satisfy
the equation
\be \ddot\varepsilon+H(2+3c^2-6w)\dot\varepsilon +
\left(\dot
H(1-3w)-15H^2w+9H^2c^2+\frac{k^2}{a^2}c^2\right)\varepsilon+\frac{k^2}{a^2}\frac{e}{\rho}=0,
\label{44}\ee where \be w={p\over \rho}  \sp c^2={\dot p\over \dot
\rho}\sp e=\delta p-c^2\delta \rho. \label{45}\ee

Experience with a similar problem, namely the cosmological perturbations in the holographic dual of the RS cosmology\footnote{Perturbations in the 5d picture of
 RS cosmology have been described in \cite{Rspert}.} \cite{hp}, indicates the following changes in the equation above:
 \begin{itemize}

 \item A modification of the time-dependent coefficients of the $\dot\varepsilon$ and $\varepsilon$ terms in (\ref{44}).
 This occurs because of the new couplings to the cosmological background.

 \item A modification of the $k$-dependence of the $\varepsilon$ and non-adiabatic terms in (\ref{44}).
 In particular, we expect the apparence of $k^4$ and $k^6$ terms originating from the quadratic curvature terms and the Cotton tensor respectively as well as
from similar terms from the scalar action (\ref{24}).
 The presence of such terms is expected to affect importantly the short distance structure of fluctuations and in particular render them invariant at early times.

 \end{itemize}

Most importantly, in line with our general discussion in the introduction, it may be that the background relevant for the study
of such perturbations may not be an approximate de Sitter background.

To qualify this we will approximately study the scalar field perturbations in the Ho\v rava-Lifshitz era.
In that case, it is the terms in the action with six derivatives that dominate.
 We will assume a flat transverse three-space for simplicity.
The quadratic part of the action of the scalar in (\ref{24})
 Fourier-transformed in transverse space and in the gauge $N=1$ reads\footnote{The observation that Ho\v rava-Lifshitz gravity
can lead to scale-invariant scalar perturbations without inflation,
 was suggested independently in \cite{muko} just before the second version of this paper
has appeared.}
\be S=\int d^3k\int dt~a^3\left[|\dot\Phi|^2+{1\over
a^6}\left(-\ell^4k^6+y_2\ell^2 k^4-y_1k^2-m^2\right)|\Phi|^2\right],
\label{46}\ee where $k\equiv \sqrt{\vec k^{\,2}}$, $\Phi(t,\vec k)$
is the Fourier transform, $\ell$ is a length scale characteristic of
the UV behavior of the scalar theory, and $y_{1,2}$ are
dimensionless coefficients. In particular, $y_1$ is the square of
the speed of light in the scalar theory. This does not have to
 be necessarily equal to that of gravitational waves in the IR.
The effective action for fluctuations $\delta\Phi$ for the scalar is
identical to (\ref{45}) with equation of motion
\be \ddot{\delta
\Phi}+3H\dot{\delta \Phi}+{\ell^4 k^6-y_2\ell^2k^4+y_1k^2+m^2\over
a^6}\delta \Phi=0. \label{47}
\ee
This suggests that at high energy
the dispersion relation is \be E^2=\ell^4{k^6\over a^6}.
\label{48}\ee Typically, a fluctuation mode oscillates if $E>>H$,
while it is frozen in the opposite limit $E<<H$. A crucial point of
standard inflation is that short distance oscillating quantum
fluctuation modes freeze fast as they become super-horizon modes.
Indeed this happens generically in the Ho\v rava-Lifshitz era for
rather general cosmological backgrounds. For this to happen, as $
{E^2\over H^2}={\ell^4~k^6\over H^2a^6}$ we should have that
$H^2a^6$ is an increasing function of time. Looking  back into the
cosmological equations (\ref{37}), we obtain that this is satisfied
for the curvature term, as well as for any matter component with
effective pressure to density ratio $w<1$. This should be contrasted
to the usual relativistic case where the condition is $a^2H^2$ being
an increasing function of time, that is violated by curvature and is
only satisfied for $w<-{1\over 3}$.

At freeze-out the physical wavelength of fluctuations should be
larger than the horizon size. Since here this is given by
$H\lambda_{phys}\simeq (H\ell)^{2\over 3}$, this implies that we
need $H\ell>>1$. This is feasible as the scale $\ell$ is not
directly constrained by low-energy data but only by precision
measurements.

The properly normalized solution to (\ref{47}) is
\be \Phi(t,\vec
k)={1\over \sqrt{2\kappa}}e^{-i\kappa\int {dt\over a^3}}\sp
\kappa\equiv \sqrt{\ell^4 k^6-y_2\ell^2k^4+y_1k^2+m^2}. \ee Going
through the standard calculation of the power spectrum  we obtain
\be \langle \delta\Phi(t,\vec k)\delta \Phi(t,\vec
k')\rangle={(2\pi)^3\over 2\kappa}\delta(\vec k+\vec k'), \ee which
leads to a scale invariant power spectrum in the UV, characterized
by the scale $\ell$. The leading classical scaling violation is set
by the coefficients of the quadratic curvature terms that may have
either sign $y_2$. However, we expect also logarithmic  scaling
violations due to UV quantum effects. They tilt the spectrum, but to
decide on which side they do, a one-loop calculation is necessary.

The fluctuations of the scalar will remain frozen until much later,
when the universe will be colder, the scalar will become first relativistic and then later non-relativistic again if massive.
Its decays to other observable fields can seed structure in the universe.
In this context, no graceful exit, neither reheating seems to be needed, and the transition is encoded in the various terms of the scalar and gravitational action that interpolate
between the Ho\v rava-Lifshitz and GR era.

There are many important issues that are left open from the qualitative discussion above.
These issues deserve further study as they suggest a different early-universe behavior from standard GR.

\vspace{.7 in}
\addcontentsline{toc}{section}{Acknowledgments}

\noindent {\bf Acknowledgements} \newline

We are grateful to P. Ho\v rava and T. Tomaras  for
valuable discussions.

 This work was  partially supported by  a European Union grant FP7-REGPOT-2008-1-CreteHEPCosmo-228644,
  an ANR grant NT05-1-41861 and a CNRS PICS grant \# 4172.

 Elias Kiritsis is on leave of absence from CPHT, Ecole Polytechnique (UMR du CNRS 7644).

\vspace{.7 in}
\addcontentsline{toc}{section}{Note Added}

\noindent {\bf Note Added} \newline

While this work was being typed, we became aware of \cite{cal} where similar issues are discussed. We seem to agree with \cite{cal} on the
regions of overlap.

After the completion of this work we also became aware of \cite{soda}. In this reference the graviton fluctuation equations are derived and studied
 around  a de Sitter spacetime, arguing for a non-zero polarisation in primordial gravitation
  waves due to the P-violating Cotton tensor dependence of the action.
We thank J. Soda for bringing his paper to our attention.

The issue of scalar perturbations and their scale invariant spectrum has been also discussed independently in \cite{muko} with similar conclusions.

\end{document}